\documentclass[aps,prl,superscriptaddress,twocolumn]{revtex4-2}

\usepackage{graphicx}
\usepackage{dcolumn}
\usepackage{bm}
\usepackage{amsmath}
\usepackage[dvipsnames]{xcolor}
\usepackage[normalem]{ulem}
\usepackage[colorlinks=true, linkcolor=blue, citecolor=blue, urlcolor=blue]{hyperref}

\begin{document}

    
\title{Correlation-Driven Spin Reorientation via Competing Anisotropy Channels in CrPS$_4$}
	
\author{Raju Baral}
\email{baralr@ornl.gov}
\affiliation{Neutron Scattering Division, Oak Ridge National Laboratory, Oak Ridge, Tennessee 37831, USA}
	
\author{Harald O. Jeschke}
\affiliation{Research Institute for Interdisciplinary Science, Okayama University, Okayama 700-8530, Japan}
	
\author{Igor I. Mazin}
\affiliation{Department of Physics and Astronomy, George Mason University, Fairfax, VA 22030, USA}
	
\author{Jue Liu}
\affiliation{Neutron Scattering Division, Oak Ridge National Laboratory, Oak Ridge, Tennessee 37831, USA}
	
\author{David Mandrus}
\affiliation{Department of Materials Science and Engineering, University of Tennessee, Knoxville, Tennessee 37996, USA}
\affiliation{Materials Science and Technology Division, Oak Ridge National Laboratory, Oak Ridge, Tennessee 37831, USA}
	
\author{Stuart Calder}
\email{caldersa@ornl.gov}
\affiliation{Neutron Scattering Division, Oak Ridge National 
    Laboratory, Oak Ridge, Tennessee 37831, USA}

\begin{abstract}
We identify a correlation-driven mechanism for the temperature-induced spin reorientation in the quasi-one-dimensional van der Waals antiferromagnet CrPS$_4$. Magnetic pair distribution function (mPDF) analysis resolves the local spin direction and shows that ferromagnetic intrachain correlations persist far above $T_\mathrm{N}$. Combining these correlations with a DFT-derived spin Hamiltonian reveals competing single-ion and exchange-anisotropy channels, with single-ion anisotropy remaining local while exchange anisotropy is renormalized as intersite correlations decay. This differential renormalization rotates the effective easy axis and captures the ordered-state canting. Above $T_\mathrm{N}$, the continued rotation beyond the model prediction delineates the limits of the dominant-chain approximation. These results establish mPDF-derived correlations as direct inputs to microscopic Hamiltonians and show how low-dimensional correlations can control magnetic anisotropy.
\end{abstract}
	
\maketitle
	
\noindent\textit{Introduction}---Spin-reorientation transitions (SRTs) occur when competing magnetic anisotropies change with temperature, field, pressure, or strain \cite{CallenCallen1966}. In conventional magnets, this balance is usually understood in terms of local crystal-field anisotropy, spin-orbit-driven magnetocrystalline anisotropy, or magnetoelastic coupling \cite{Chikazumi1997,Coey2010}. In low-dimensional magnets, where anisotropy stabilizes magnetic order against enhanced thermal fluctuations \cite{MerminWagner1966,BanderMills1988,Gibertini2019}, a distinct route can arise in which exchange anisotropy is renormalized by the decay of intersite spin correlations, whereas single-ion anisotropy remains local \cite{RozsaAtxitia2023}. Directly identifying this correlation-dependent renormalization requires a probe that simultaneously measures local spin correlations and spin direction, yet such a combined experimental and microscopic demonstration has not been established.
	
\begin{figure}[b]
\centering
\includegraphics[width=0.8\linewidth]{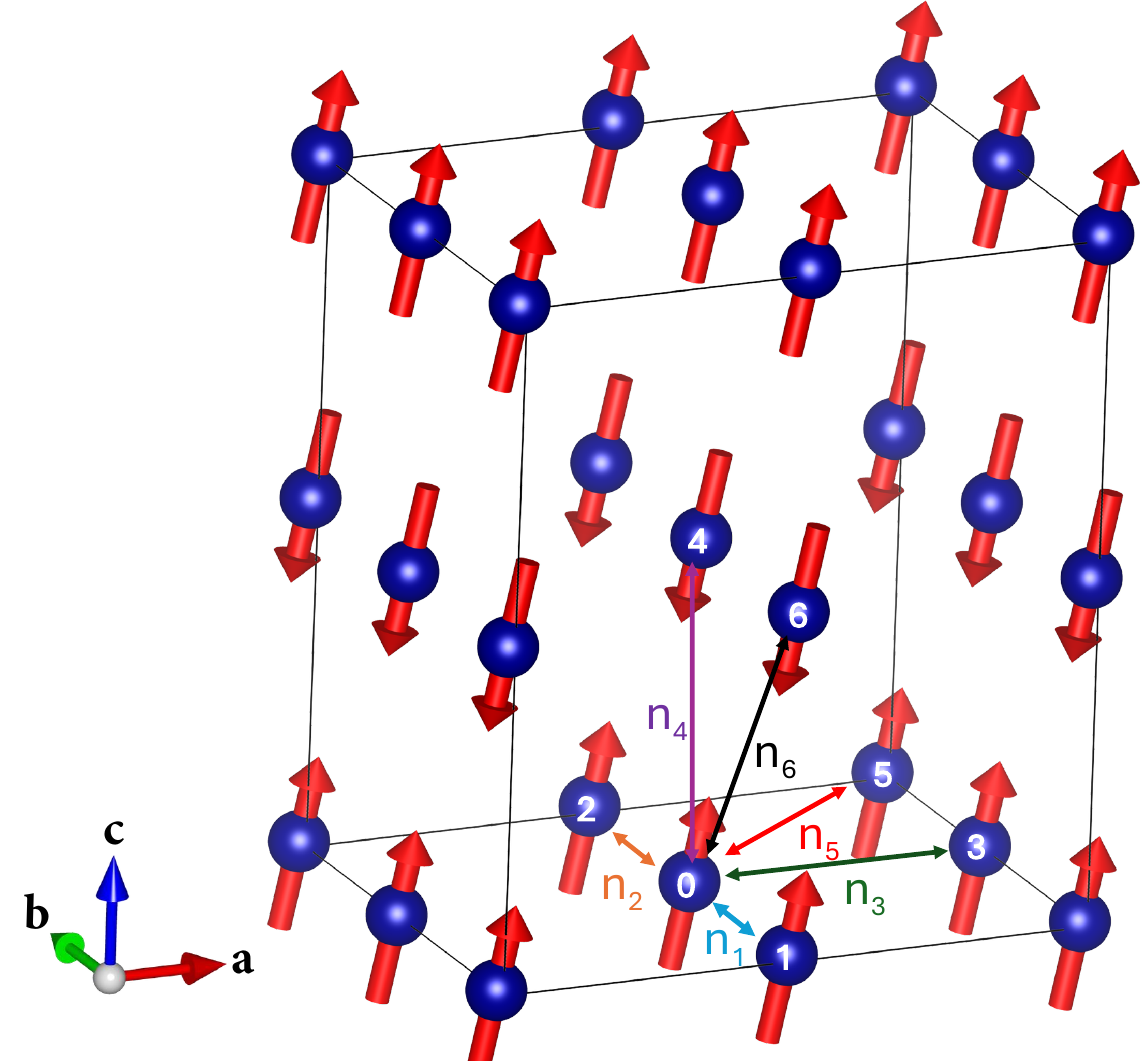}
\caption{\label{fig:magStructure} 
Magnetic structure of CrPS$_4$, where the Cr spins are canted by 12$^\circ$ from the $c$ axis at 2 K. The six nearest-neighbor Cr–Cr shells ($n1$–$n6$) used in the correlation analysis are labeled, with $n1$,$n2$ forming the dominant intrachain bonds.}
\end{figure}

CrPS$_4$ provides a direct platform for testing this mechanism. It combines low-dimensional magnetic correlations, weak anisotropy, and a temperature-dependent spin direction in an air-stable van der Waals antiferromagnetic semiconductor. CrPS$_4$ remains magnetic down to exfoliated monolayers \cite{son2021air,lee2017structural} and has attracted interest for magnon transport \cite{Shaomian_CPS} and tunnel junction devices \cite{PhysRevApplied.16.024011,Li_CPS,Fan_CPS}. Below $T_\mathrm{N}=37$~K, the Cr$^{3+}$ moments form an A-type antiferromagnetic structure, with ferromagnetic alignment within the $ab$-plane and antiferromagnetic stacking along $c$ [Fig.~\ref{fig:magStructure}] \cite{calder2020mag,CPS_Peng}. Strong quasi-one-dimensional (1D) Cr--Cr chains dominate over weaker in-plane and interlayer couplings \cite{calder2020mag}, producing persistent short-range correlations and an anisotropy balance sensitive to thermal fluctuations.

CrPS$_4$ exhibits several reorientation phenomena that point to a delicate anisotropy balance, including an unusual two-stage magnetic transition \cite{calder2020mag}, metamagnetic transitions beginning at fields as low as 0.7~T \cite{pei2016,calder2020mag,CPS_Peng}, a pressure-driven SRT \cite{Peng_Pressure_CPS}, and torque anomalies reflecting weak magnetocrystalline anisotropy \cite{Seo2024_CrPS4}. These observations establish CrPS$_4$ as a model system in which small changes to correlations, lattice geometry, or magnetic field can rotate the preferred spin direction.

Here we combine magnetic PDF analysis \cite{frandsen2015mPDFMnO,frandsen2022diffpympdf} of neutron total scattering with DFT+U-derived spin Hamiltonians to show that the temperature-driven SRT in CrPS$_4$ arises from the differential renormalization of single-ion and exchange-anisotropy channels by real-space spin correlations, naturally accounting for the two-stage character of the transition.
	
\begin{figure*}[tbp]
\centering
\includegraphics[width=150mm]{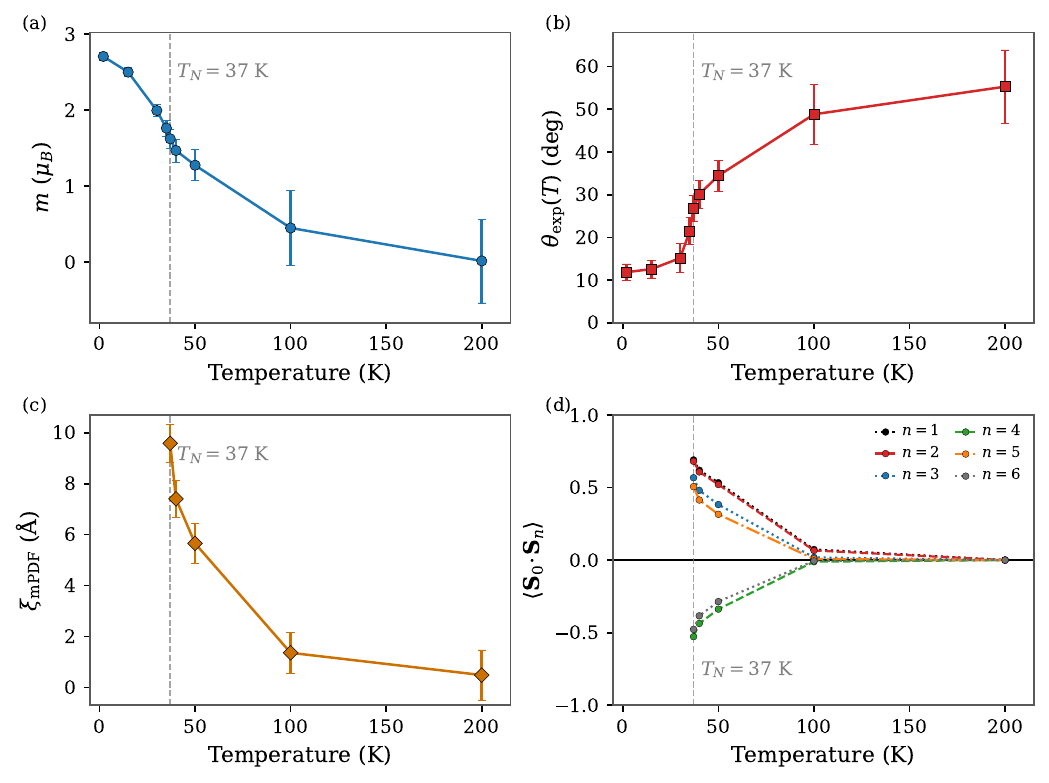}
\caption{\label{fig:m_th_xi} 
Temperature evolution of the (a) local magnetic order parameter, (b) spin angle, (c) correlation length for $T > T_\mathrm{N}$, and (d) shell-resolved spin-spin correlation function, $\langle \mathbf{S}_0 \cdot \mathbf{S}_n \rangle$, of the first six nearest neighbors for $T > T_\mathrm{N}$. Positive values correspond to net ferromagnetic (FM) and negative to antiferromagnetic (AFM) correlations. The persistence of positive $n=1,2$ correlations to 100 K demonstrates that the paramagnetic state remains locally chain-like even after the collapse of long-range correlations.}
\end{figure*}

\noindent\textit{Local magnetic correlations}---Temperature-dependent total scattering measurements were performed on NOMAD \cite{NOMAD}, with field-dependent data collected on HB-2A \cite{mPDF_HB2A_Baral}. mPDF analysis over 2--300 K and 1.5--30~\AA\ yield the local magnetic order parameter, spin direction, and correlation length $\xi_{\mathrm{mPDF}}$. The local moment [Fig.~\ref{fig:m_th_xi}(a)] decreases gradually with temperature, while the polar angle $\theta$ [Fig.~\ref{fig:m_th_xi}(b)] increases from $12^\circ$ at 2~K to $58^\circ$ at 200~K. The best-fit azimuthal angle remains near the $a$ direction. The refined rotation is stable against fitting-range and correlation-length tests described in the Supplemental Material. The paramagnetic correlation length [Fig.~\ref{fig:m_th_xi}(c)] decreases with temperature and becomes indistinguishable from zero by 200~K.
	
The spin correlation function $\langle \mathbf{S}_0 \cdot \mathbf{S}_n \rangle$ [Fig.~\ref{fig:m_th_xi}(d)] shows that the chain neighbors $n=1,2$ remain ferromagnetically correlated to 100 K. The interchain in-plane shell $n=3$ is also ferromagnetic at 50 K, whereas several more distant out-of-plane shells are weakly antiferromagnetic. The slower decay of the $n=1,2$ correlations provides direct real-space evidence for persistent intrachain FM correlations above $T_\mathrm{N}$ and motivates the correlation-renormalization treatment below.

Field-dependent mPDF analysis further show that the local correlations evolve from predominantly AFM at low field to FM by 6 T, consistent with the weak anisotropy balance inferred from single-crystal spin-flop and spin-flip studies for $H\parallel c$ (Supplemental Material) \cite{calder2020mag,CPS_Peng}.
	
\noindent\textit{Correlation-renormalized anisotropy}---Having established the quasi-1D correlation structure from the mPDF data (Fig.~\ref{fig:m_th_xi}), we now construct the dominant-chain anisotropy model. 	The key observation is that the intrachain correlations ($n=1,2$) remain positive to 100 K, even though the fitted mPDF
correlation length has shortened by 50 K to only a few \AA, comparable to the nearest-neighbor Cr--Cr spacings. Thus, the paramagnetic state does not retain extended correlations, but it does preserve a local 1D correlation pattern on the first chain bonds.

The microscopic Hamiltonian for the dominant chain bonds can be written as
\begin{equation}
\label{eq:spin_hamiltonian}
\begin{aligned}
\mathcal{H}_{ij} =\;& J\,\mathbf{S}_i\cdot\mathbf{S}_j
			+ J_x S_{ix}S_{jx} + J_y S_{iy}S_{jy} \\
			&+ J_{xz}(S_{ix}S_{jz}+S_{iz}S_{jx}) \\
			&+ K_y(S_{iy}^2+S_{jy}^2) + K_z(S_{iz}^2+S_{jz}^2) \\
			&+ K_{xz}(S_{ix}S_{iz}+S_{jx}S_{jz}) \, .
\end{aligned}
\end{equation}
Here $y\parallel b$, $x\parallel a$, $z\approx c$, and $\mathbf{S}_i$ are spin-$3/2$ operators for Cr$^{3+}$. These are the only anisotropic terms allowed by symmetry on the $n{=}1,2$ intrachain bonds (Supplemental Material). Antisymmetric Dzyaloshinskii--Moriya exchange is absent on both chain bonds since each bond center is an inversion center of the $C2/m$ structure. We apply this Hamiltonian in a dominant-chain approximation, retaining the anisotropy terms on the strongly ferromagnetic intrachain bonds $J_1$ and $J_2$ \cite{calder2020mag}. For fully correlated chains, the effective anisotropies are $D_y=K_y+J_y$, $D_z=K_z+J_z$, and $D_{xz}=K_{xz}+J_{xz}$, where the off-diagonal term $D_{xz}$ cants the spins away from the high-symmetry axes. At finite temperature, the exchange-anisotropy terms are renormalized by the intersite correlation factor according to $D_\mu(T)=K_\mu+\xi(T)J_\mu$, with $\mu=y,z,xz$. Thus, decaying intersite correlations suppress the exchange-anisotropy contribution while leaving the on-site $K$ terms local, rotating the preferred spin direction.
	
\begin{figure*}[tb]
\centering
\includegraphics[width=\textwidth]{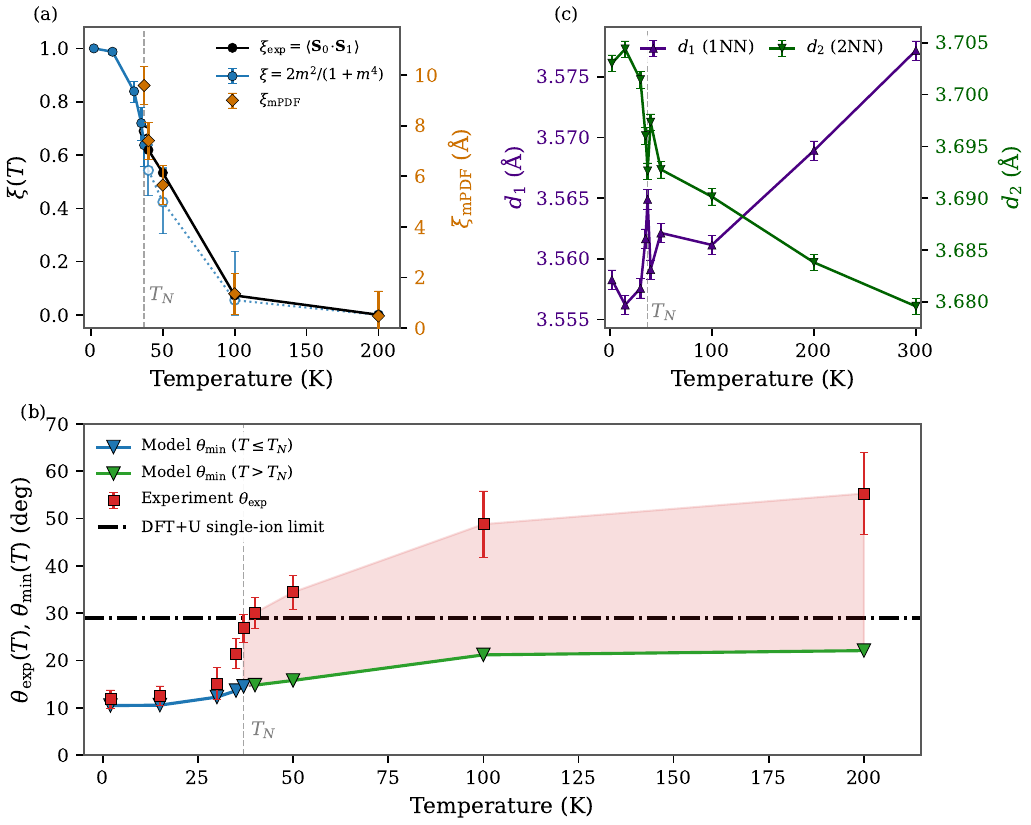}
\caption{\label{fig:mechanism}
(a) Correlation-renormalization factor $\xi(T)$ from the moment-based estimate $\xi = 2m^2/(1+m^4)$ (blue circles, filled for $T \leq T_\mathrm{N}$, open above $T_\mathrm{N}$) and the directly measured intrachain correlation	$\xi_{\mathrm{exp}}=\langle \mathbf{S}_0 \!\cdot\! \mathbf{S}_1\rangle$ from mPDF analysis (used as model input for $T>T_\mathrm{N}$). The mPDF correlation length $\xi_{\mathrm{mPDF}}$ is shown on the right axis. Agreement near $T_\mathrm{N}$ supports the use of the Gaussian estimator in the ordered phase. (b) Model-predicted equilibrium spin angle $\theta_{\min}(T)$ (triangles) compared with the experimentally refined spin angle $\theta_{\mathrm{exp}}(T)$. Below $T_\mathrm{N}$, $\theta_{\min}$ is computed using the moment-based $\xi$ and above $T_\mathrm{N}$ it is computed using the directly measured $\xi_{\mathrm{exp}}$. The dashed horizontal line marks the DFT+U single-ion estimate obtained from the ratio of the single-ion anisotropy terms $K_{xz}$ and $K_z$, corresponding to the limit where exchange-anisotropy terms are suppressed. The model captures the ordered-state canting quantitatively, while above $T_\mathrm{N}$ the experimental rotation continues beyond this simplified single-chain description. The shaded region marks the additional rotation not captured by the simplified model, reflecting the approximations discussed in the text. (c) Nearest-neighbor ($d_1$) and next-nearest-neighbor ($d_2$) Cr--Cr chain distances from the PDF analysis, both showing anomalies at $T_\mathrm{N}$ consistent with spin--lattice coupling.}
\end{figure*}

To capture the thermal evolution of the anisotropy, we introduce a correlation-renormalization factor $\xi(T)\equiv \langle S_\alpha S'_\alpha\rangle/\langle S_\alpha^2\rangle$, the ratio of intersite to on-site spin correlations along a chain. Within the ordered phase, where shell-resolved correlations are not separately accessible from the mPDF data, we estimate $\xi(T)$ from the temperature-dependent local moment using
\begin{equation}
	\xi(T)=\frac{2m(T)^2}{1+m(T)^4},
\end{equation}
where $m(T)$ is the local moment normalized to its low-temperature value. This expression follows from treating the reduction of the local moment as arising from Gaussian angular fluctuations of locally saturated spins, as derived in the Supplemental Material. This estimate treats each spin as fluctuating independently about a common easy axis, which is appropriate in the ordered phase where the sublattice magnetization is well defined but individual pair correlations cannot be isolated from the mPDF analysis. Above $T_\mathrm{N}$, where long-range order is lost, we use the directly measured intrachain correlation $\xi_{\mathrm{exp}}(T) \equiv \langle \mathbf{S}_0\!\cdot\!\mathbf{S}_1\rangle$, i.e.\ the $n{=}1$ value plotted in Fig.~\ref{fig:m_th_xi}(d) (spins are normalized to unit length, so $\langle S^2\rangle = 1$). The two estimators agree quantitatively in the vicinity of $T_\mathrm{N}$ [Fig.~\ref{fig:mechanism}(a)], providing an internal consistency check on the Gaussian approximation, and together yield a model-independent determination of $\xi(T)$ on either side of the transition. As $\xi(T)$ decreases, only the exchange-anisotropy ($J$) contributions in the effective Hamiltonian are renormalized. The single-ion ($K$) terms, set primarily by the local Cr crystal-field environment, are not directly renormalized by intersite correlations. This differential renormalization of $J$ versus $K$ is the central physical mechanism driving the SRT.
	
For the observed rotation within the $ac$ plane, the relevant anisotropy parameters are $K_z = -0.35$\,K, $J_z = -0.12$\,K, $K_{xz} = -0.17$\,K, and $J_{xz} = +0.08$\,K for the dominant chain bonds, as obtained from DFT+U energy mapping at $U = 2$\,eV (Supplemental Material). The single-ion-only estimate, set by $K_{xz}/K_z$, provides the DFT+U single-ion reference shown as the dashed horizontal line in Fig.~\ref{fig:mechanism}(b). The same $U=2$\,eV parametrization also reproduces the experimental N\'eel temperature in classical Monte Carlo simulations (Supplemental Material), providing an internal consistency check on the exchange energy scale. The ratio $|D_{xz}/D_z|\sim 0.2$ produces a small but finite canting of the easy axis from $c$ within the $ac$ plane, consistent with the equilibrium spin tilt of ${\sim}12^\circ$ measured at low temperature.

Combining the DFT+U-derived anisotropy with the experimentally determined $\xi(T)$ produces the results shown in Fig.~\ref{fig:mechanism}(b). Within the ordered phase, the dominant-chain model reproduces the canting quantitatively. The predicted equilibrium angles of $\theta_{\min} = 10.5^\circ$ at 2~K and $12.3^\circ$ at 30~K agree with the experimental values of $12\pm 2^\circ$ and $15\pm 3^\circ$ within the measurement uncertainty. This agreement is obtained with no additional adjustable parameters because the anisotropic exchange is computed ab initio and $\xi(T)$ is determined experimentally.
	
The structural anomalies at $T_\mathrm{N}$ acquire a microscopic interpretation in this framework. As shown in Fig.~\ref{fig:mechanism}(c), the Cr--Cr chain distances exhibit pronounced and opposite anomalies at $T_\mathrm{N}$, with the $d_1$ distance showing an upturn and $d_2$ showing a sharp dip. These coincident structural and magnetic anomalies indicate spin--lattice coupling through the Cr--S--Cr superexchange pathways. The dominant chain exchanges arise from near-90$^\circ$ Cr--S--Cr superexchange (Cr--S--Cr angles of $94^\circ$ and $97^\circ$ in the $C2/m$ structure \cite{calder2020mag}), where the Goodenough--Kanamori exchange depends sensitively on bond geometry. The ${\sim}0.15\%$ anomaly in the $J_1$ Cr--Cr distance and the ${\sim}0.13\%$ contraction of the $J_2$ distance at $T_\mathrm{N}$ act in opposite senses and are expected to provide a secondary modulation of the anisotropic chain couplings.
	
At 100~K, where $\xi$ is directly measured rather than estimated ($\xi_{\mathrm{exp}}\approx 0.07$), the chain-only model predicts $\theta\approx 21^\circ$ while experiment shows ${\sim}49^\circ$. 	The model is deliberately simplified, it neglects the compass terms $J_{xy}$ and $K_{xy}$ allowed by the phosphorus-induced symmetry lowering, encodes the suppression of all exchange-anisotropy channels through a single correlation parameter $\xi(T)$, and relies on DFT+U-derived anisotropy parameters that are intrinsically small (${\lesssim}\,0.35$~K), may themselves have some temperature dependence, and carry only semiquantitative accuracy at this energy scale. That even this oversimplified treatment reproduces the correct sense and scale of the spin rotation, quantitatively in the ordered phase and qualitatively above $T_\mathrm{N}$, confirms that the differential renormalization of exchange versus single-ion anisotropy is the essential mechanism driving the SRT.

\noindent\textit{Conclusion}---By combining mPDF-derived real-space correlations with a DFT+U-derived spin Hamiltonian, we have demonstrated that the SRT in CrPS$_4$ arises from the differential thermal renormalization of exchange and single-ion anisotropy channels, with no additional adjustable parameters in the ordered phase. Even an intentionally simplified dominant-chain model captures the essential physics of the rotation, confirming that the mechanism is robust. Cr--Cr distance anomalies at $T_\mathrm{N}$ further support spin--lattice coupling as an additional tuning parameter of the same anisotropy balance. The essential ingredients of this mechanism, competing single-ion and exchange anisotropy on bonds with strong correlation anisotropy, are likely present in other low-dimensional van der Waals antiferromagnets. CrSBr, which shares a quasi-1D ferromagnetic Cr-chain topology \cite{Liebich2025} and itself exhibits a layer-dependent spin reorientation \cite{YeACSNano2022}, is a natural candidate for testing this framework, while the MPX$_3$ family ($M=$ Mn, Fe, Ni and $X=\mathrm{S}$, Se) offers systematic chemical control of magnetic anisotropy and correlation dimensionality \cite{Burch2018}. These results establish that in low-dimensional magnets the local pattern of spin correlations, not merely their average, can govern magnetic anisotropy, and that mPDF analysis provides the real-space inputs needed to connect microscopic Hamiltonians to observed spin reorientation.

\noindent\textit{Acknowledgments}---This research used resources at the High Flux Isotope Reactor and Spallation Neutron Source, a DOE Office of Science User Facility operated by Oak Ridge National Laboratory. Beam time was allocated on HB-2A under proposal IPTS-33026 and NOMAD under proposal IPTS-28185. H.O.J. acknowledges support through JSPS KAKENHI Grants No. 24H01668 and No. 25K08460. Part of the computation in this work has been done using the facilities of the Supercomputer Center, the Institute for Solid State Physics, the University of Tokyo. I.I.M. acknowledges support from the National Science Foundation under Award No. DMR-2403804. This research also used beamline BXDS-WHE of the Brockhouse Diffraction Sector at Canadian Light Source. 

	
\bibliography{ref}
	
\end{document}


\title{Supplemental Information: Correlation-Driven Spin Reorientation via Competing Anisotropy Channels in CrPS$_4$}

\author{Raju Baral}
\email{baralr@ornl.gov}
\affiliation{Neutron Scattering Division, Oak Ridge National Laboratory, Oak Ridge, Tennessee 37831, USA}
	
\author{Harald O. Jeschke}
\affiliation{Research Institute for Interdisciplinary Science, Okayama University, Okayama 700-8530, Japan}
	
\author{Igor I. Mazin}
\affiliation{Department of Physics and Astronomy, George Mason University, Fairfax, VA 22030, USA}
	
\author{Jue Liu}
\affiliation{Neutron Scattering Division, Oak Ridge National Laboratory, Oak Ridge, Tennessee 37831, USA}
	
\author{David Mandrus}
\affiliation{Department of Materials Science and Engineering, University of Tennessee, Knoxville, Tennessee 37996, USA}
\affiliation{Materials Science and Technology Division, Oak Ridge National Laboratory, Oak Ridge, Tennessee 37831, USA}
	
\author{Stuart Calder}
\email{caldersa@ornl.gov}
\affiliation{Neutron Scattering Division, Oak Ridge National Laboratory, Oak Ridge, Tennessee 37831, USA}


\maketitle

\section{Experimental methods}
\label{sec:exp}

\subsection{Sample synthesis}

Sample synthesis and characterization of CrPS$_4$ is described in Ref.~\cite{calder2020mag}.

\subsection{Neutron scattering measurements}
  
Total scattering experiments were carried out at NOMAD, located at the Spallation Neutron Source (SNS), Oak Ridge National Laboratory (ORNL) \cite{NOMAD}. CrPS$_4$ powder was loaded into a 6 mm vanadium can and data were collected from 2--300 K. The automatic data reduction scripts at NOMAD were used to obtain real-space pair distribution functions by Fourier transformation with $Q_{\mathrm{max}}=25~\mathrm{\AA^{-1}}$. Magnetic-field-dependent neutron scattering data were collected on the same sample on the HB-2A powder diffractometer \cite{mPDF_HB2A_Baral,MnPS3_Se3_mPDF,mPDF_HB2A_MST,mPDF_HB2A_ZnMnTe} at the High Flux Isotope Reactor (HFIR) at from 5--65 K under applied fields of 0--6 T. For the field measurements, the powder sample was converted into 6 mm pellets, to avoid grain reorientation. 

\subsection{X-ray scattering measurements}

An X-ray total scattering experiment was performed at beamline BXDS-WHE of the Brockhouse Diffraction Sector at Canadian Light Source. The sample was loaded into polyimide capillaries, sealed, and measured at 80, 200 and 300 K. A maximum momentum transfer of 25~\AA$^{-1}$ was used in the Fourier transform to obtain the x-ray PDF, following the processing methodology in Ref.~\onlinecite{Burns:xx5017}. 

\section{Analysis of total scattering data: atomic and magnetic PDF fits}
\label{sec:pdf}

\subsection{Neutron and X-ray PDF analysis at 300 K}
\begin{figure}[htbp]
    \centering
    \includegraphics[width=0.85\linewidth]{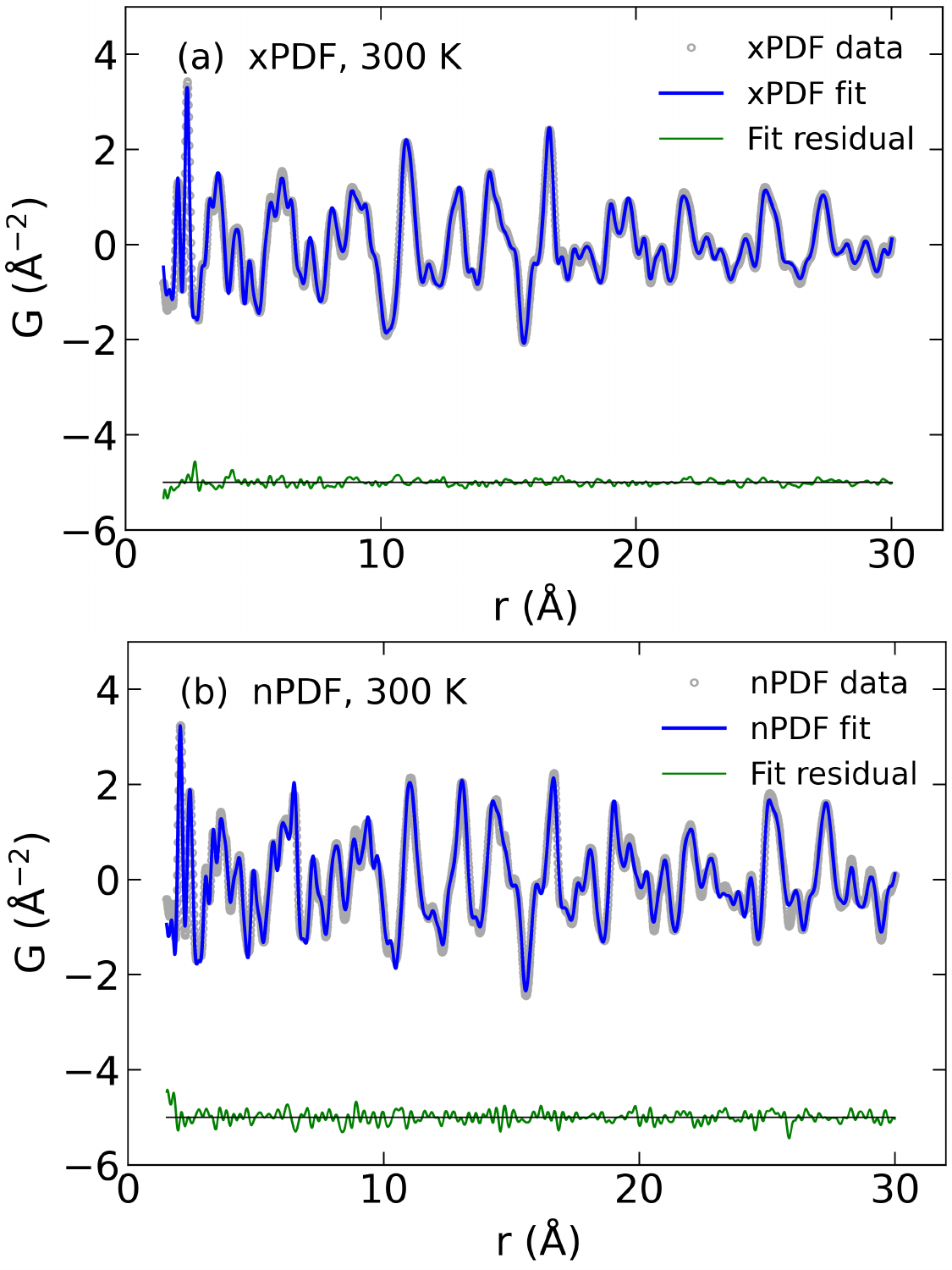}
    \caption{\label{fig:xPDF} 
        (a) X-ray PDF fit of CrPS$_4$ at 300 K using a fitting range of 1.5\,--\,30\,\AA. (b) Same as (a) but for neutron PDF data.
    }
\end{figure}

Fig.~\ref{fig:xPDF} shows X-ray and neutron PDF fits to the 300 K data over 1.5\,--\,30\,\AA, confirming that the refined room-temperature structure describes both datasets.

\subsection{Magnetic pair distribution function analysis}

\begin{figure}[htbp]
	\centering
	\includegraphics[width=0.8\columnwidth]{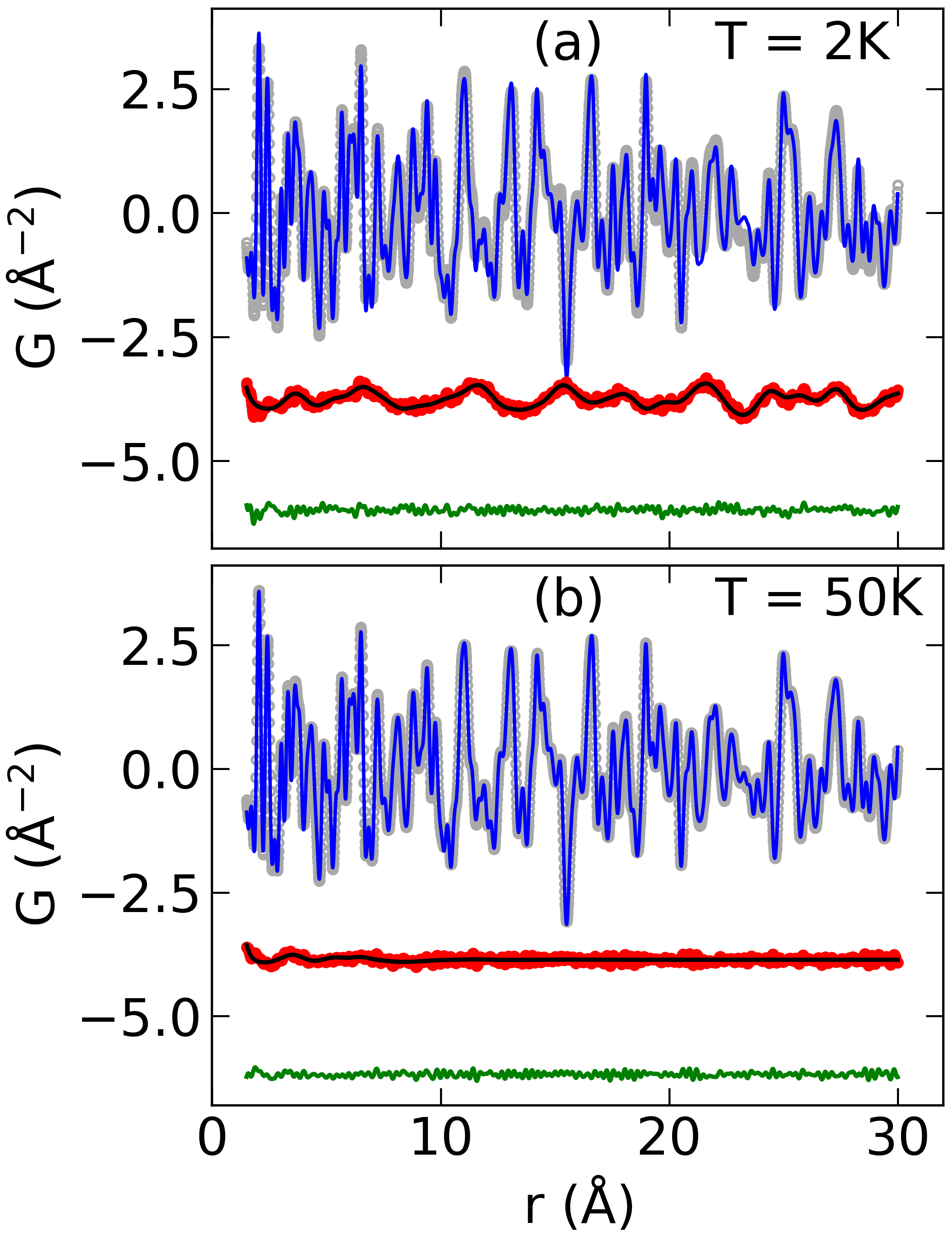}
	\caption{\label{fig:totalPDF_2K} 
		(a) Atomic and magnetic PDF fit of CrPS$_4$ at 2 K using a fitting range of 1.5\,--\,30\,\AA. 
		The gray and blue symbols represent the total PDF data and total PDF fit. 
		Note that the total PDF fit is the sum of atomic and magnetic PDF fits. 
		The red, black and green curves correspond to the isolated magnetic PDF data, magnetic PDF fit and overall fit residual, respectively. (b) Same as (a), but for 50 K, above $T_\mathrm{N}$.
	}
\end{figure}

Structural fits were performed with PDFgui \cite{Farrow2007PDFgui} over 1.5\,--\,30\,\AA~using the published CrPS$_4$ structure. mPDF analyses used \textit{diffpy.mpdf}~\cite{frandsen2022diffpympdf}. At 2 K [Fig.~\ref{fig:totalPDF_2K}(a)], the refined spin direction is $\theta = 12 \pm 2^\circ$ from the $c$ axis and $\phi = 0^\circ$ from the $a$ axis, with local magnetic order $m = 2.70 \pm 0.05\,\mu_\mathrm{B}$, consistent with prior neutron diffraction results \cite{calder2020mag}. At 50 K [Fig.~\ref{fig:totalPDF_2K}(b)], above $T_\mathrm{N}$, reduced but detectable mPDF features indicate short-range magnetic correlations. For visualization, a temperature-averaged residual associated with preferred orientation or texture was subtracted after refinement, as discussed below.

\subsection{Texture correction on mPDF data}
\begin{figure}[htbp]
    \centering
    \includegraphics[width=\linewidth]{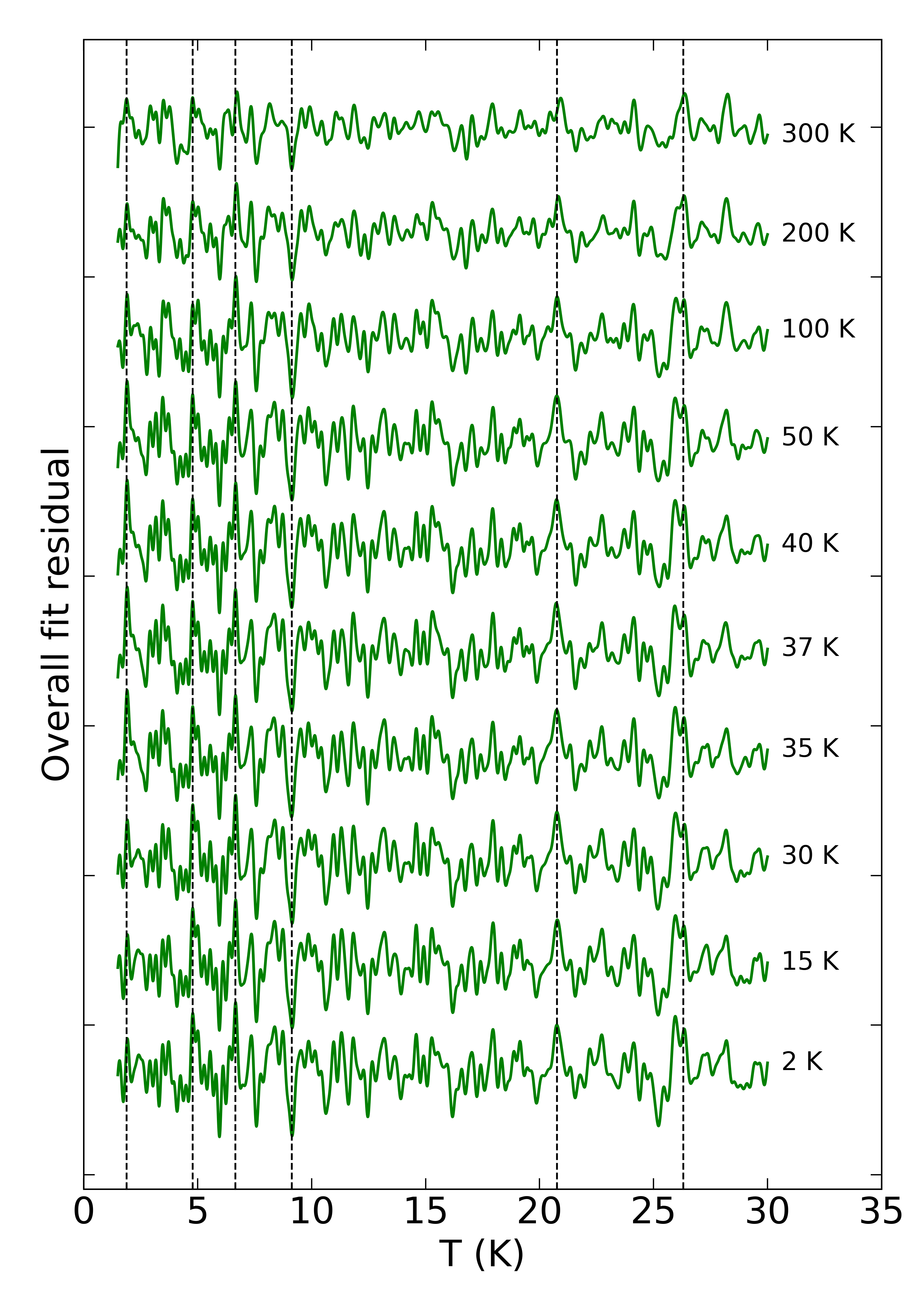}
    \caption{\label{fig:overallFitRes} 
Temperature dependence of the overall fit residual. The black dashed lines highlight some sharp features that persist across all temperatures.
    }
\end{figure}

Inspection of the overall fit residuals from the atomic and magnetic PDF refinements reveals the presence of sharp oscillatory features (see Fig.~\ref{fig:overallFitRes}). These features may arise from Fourier-transform artifacts or, more likely, sample texture/preferred orientation. At present, no well-established method exists for accounting for preferred-orientation effects within PDF refinements. Therefore, only for visualization purposes, we subtract the temperature-averaged overall fit residual from each mPDF dataset. This correction is applied only after performing the magnetic PDF refinement and is used solely to improve the clarity of the plotted mPDF patterns.

\subsection{Azimuthal-angle dependence in the mPDF fits}
\begin{figure}[htbp]
    \centering
    \includegraphics[width=\columnwidth]{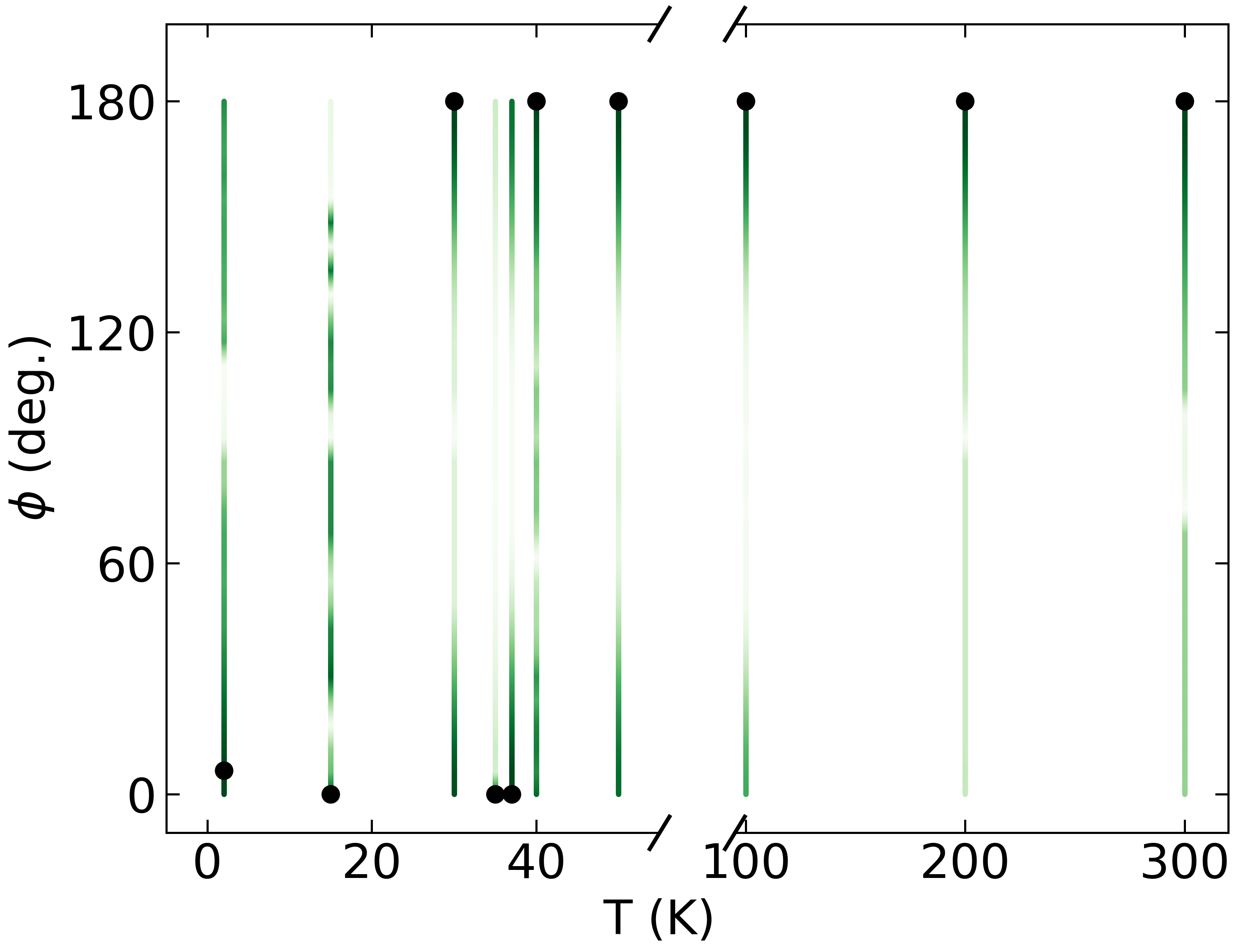}
    \caption{\label{fig:phiDependency} 
Variation of azimuthal angle, $\phi$ (angle from $a$ axis) as a function of temperature.
    }
\end{figure}

To evaluate the sensitivity of the azimuthal angle in the mPDF refinements, we performed a series of mPDF fits at each temperature by varying the azimuthal angle (defined with respect to the $a$ axis) from 0$^{\circ}$ to 180$^{\circ}$ in increments of 5$^{\circ}$. The goodness-of-fit metric $R_w$ was computed at each temperature and the angle yielding the lowest $R_w$  was identified as the best-fit azimuthal angle. These best-fit angles are indicated by the black circles, while the green shaded vertical bars represent the confidence interval associated with each best-fit value. The shading progressively lightens as $R_w$ increases away from its minimum, fading to white once the value exceeds the minimum $R_w$ for that temperature by 0.5\% or more. A threshold of 0.5\% was chosen to mark a meaningful change in the quality of fit. The resulting azimuthal angles show limited temperature dependence, with the fits consistently favoring orientations near 0$^{\circ}$ or 180$^{\circ}$ relative to the $a$ axis.

\subsection{Field-dependent mPDF fits}

\begin{figure*}[tb]
	\centering
	\includegraphics[width=0.8\textwidth]{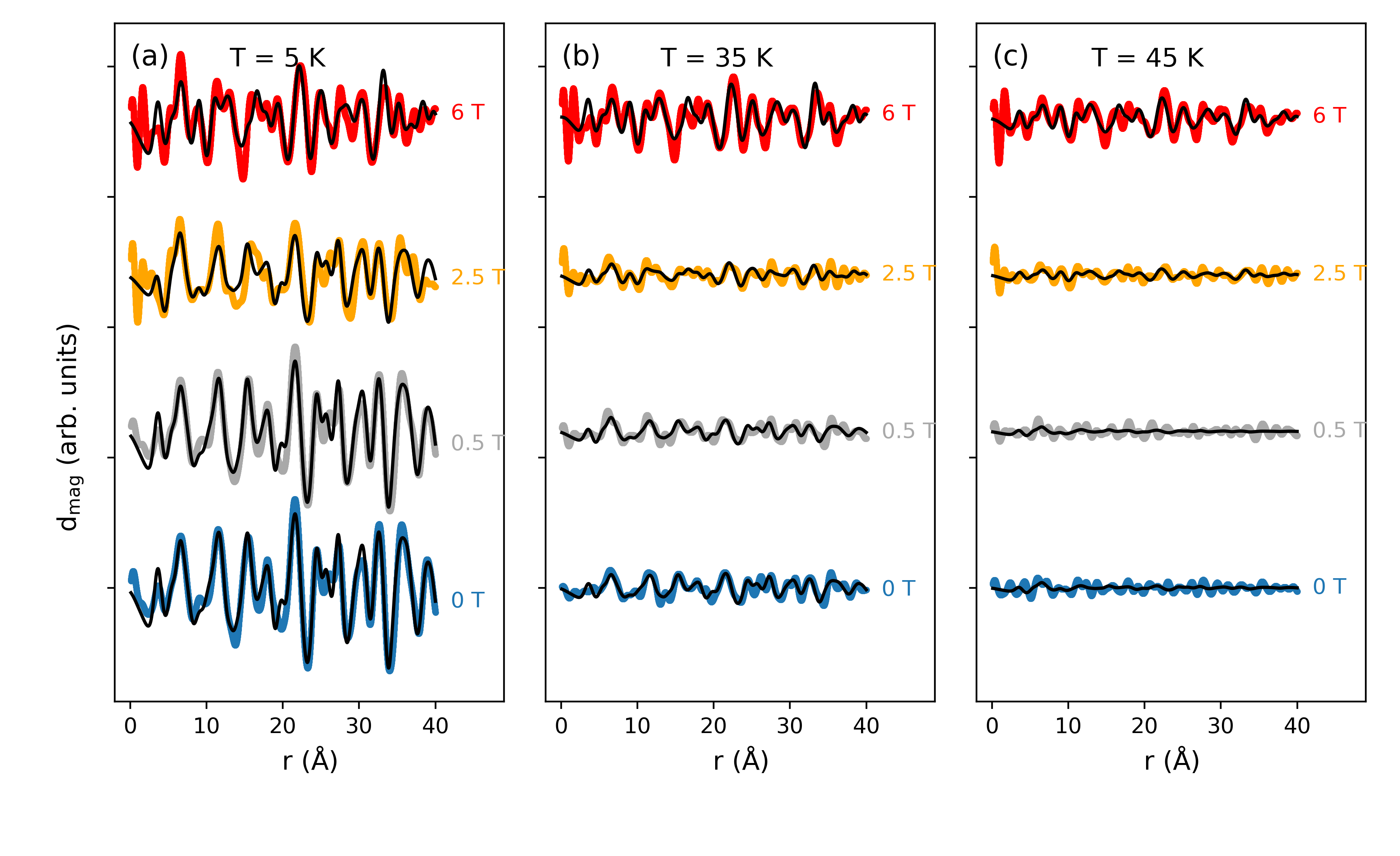}
	\caption{\label{fig:field_mPDF} 
		(a) Field-dependent magnetic pair distribution functions (mPDFs) measured at 5 K under applied magnetic fields of 0, 0.5, 2.5 and 6 T. (b) and (c) show the corresponding field-dependent mPDFs measured at 35 and 45 K, respectively.
	}		
\end{figure*}

The correlation-renormalization picture predicts that the magnetic ground state should be particularly sensitive to perturbations when intersite correlations are weak. To test this, we extended the mPDF measurements to applied magnetic fields. Single-crystal studies have shown spin-flop and spin-flip metamagnetic transitions in CrPS$_4$ for $\mathbf{H}\parallel c$ \cite{calder2020mag,CPS_Peng,pei2016}, in powders these transitions appear as phase coexistence \cite{CPS_Peng}.

Magnetic field--dependent neutron scattering measurements were performed at 5, 35 and 45~K under applied fields of 0, 0.5, 2.5 and 6~T. The 65~K zero-field data were used as a reference to isolate the magnetic scattering by subtracting the nuclear contribution. The resulting magnetic scattering patterns were Fourier transformed with $Q_{\mathrm{max}}=6~\mathrm{\AA^{-1}}$ to obtain the mPDF data.

The field-dependent mPDF refinements used a mixed magnetic PDF model with two propagation vectors of $k_1=(0,0,0.5)$ for antiferromagnetic (AFM) correlations and $k_2=(0,0,0)$ for ferromagnetic (FM) correlations. For the zero-field data, only $k_1=(0,0,0.5)$ was used, consistent with the zero-field magnetic structure.

At all three temperatures, the fitted local correlations evolve from AFM-like at low fields to predominantly FM at 6~T (Fig.~\ref{fig:field_mPDF} and Table~\ref{tab:AF_FM_components}). At 5~K [Fig.~\ref{fig:field_mPDF}(a)], the overall mPDF intensity decreases with increasing field, reflecting the progressive reorientation from the AFM configuration toward a field-aligned FM state. At intermediate fields, especially 2.5~T, the refinements reveal coexistence of AFM and FM components, consistent with spin-flop behavior and phase mixing within the powder sample. At 6~T, the FM component dominates.

\begin{figure}[htbp]
	\centering
	\includegraphics[width=\linewidth]{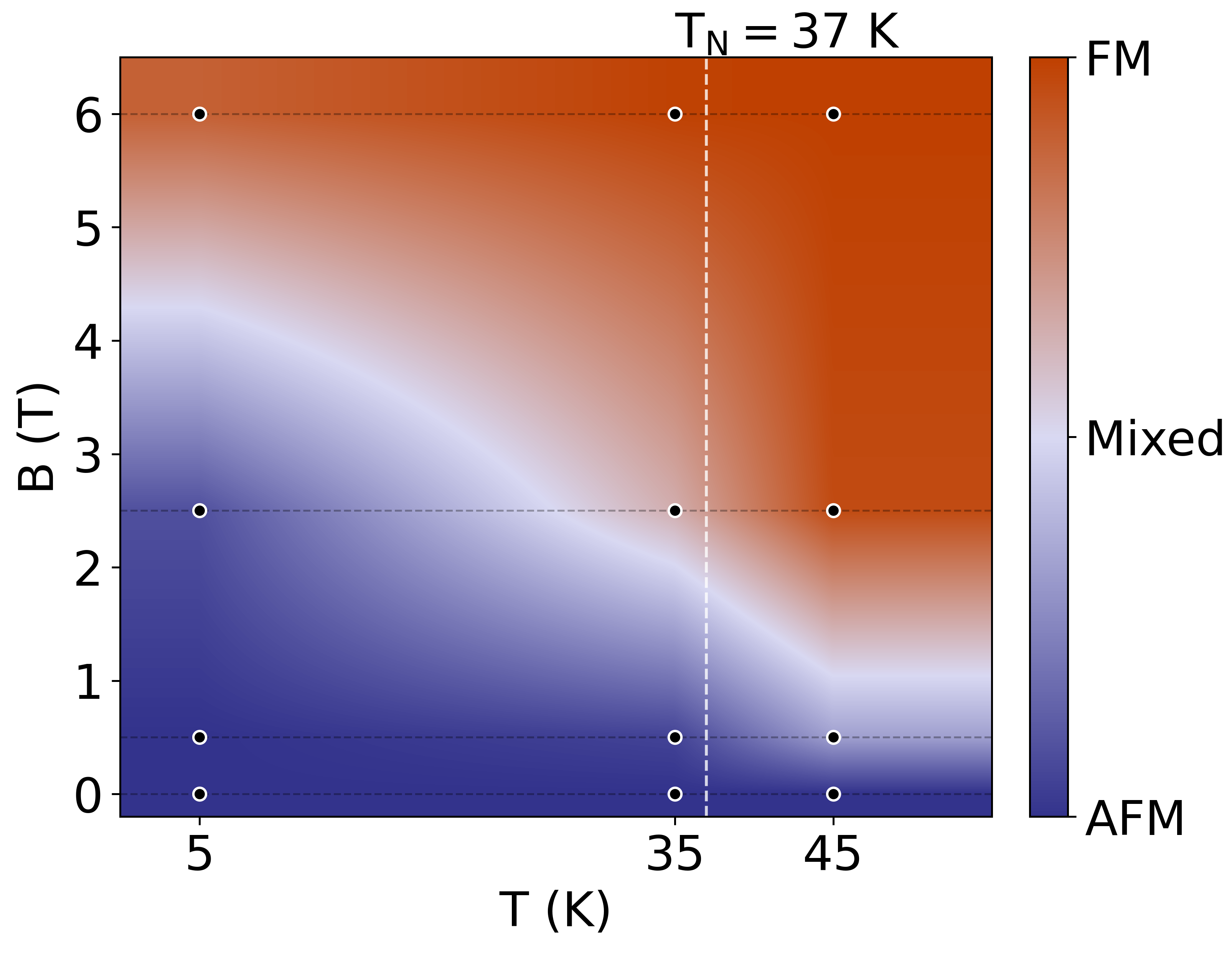}
	\caption{\label{fig:phase_diagram} 
		Field-temperature map of local AFM and FM correlation fractions from field-dependent mPDF measurements. The color scale represents the relative fraction of ferromagnetic (FM) and antiferromagnetic (AFM) correlations. The dashed vertical line marks the N\'eel temperature ($T_\mathrm{N}=37$\,K). Note: The plotted data correspond to measurements performed at 5, 35 and 45~K. The color gradient between these points is interpolated for visualization. The quantitative interpretation is reliable primarily in the vicinity of the experimental temperatures. 
	}
\end{figure}

The same trend persists near and above $T_\mathrm{N}$. At 35~K, the FM fraction grows rapidly with field, while at 45~K [Fig.~\ref{fig:field_mPDF}(c)] a weak zero-field magnetic signal indicates short-range AFM-like correlations above the long-range ordering temperature. These correlations are rapidly suppressed by field and replaced by FM alignment, consistent with reduced intersite correlations in the quasi-1D regime. The field-temperature map in Fig.~\ref{fig:phase_diagram} summarizes this evolution and shows that the local correlation network is readily tuned by modest fields, as expected when competing anisotropy channels have comparable energy.

\begin{table}[h!]
	\caption{Antiferromagnetic and ferromagnetic components at different temperatures and magnetic fields.}
	\label{tab:AF_FM_components}
	\begin{ruledtabular}
		\renewcommand{\arraystretch}{1.3}
		\begin{tabular}{c c c c}
			T (K) & Magnetic Field  & AF component  & FM component\\
             &  (T) & (\%) & (\%) \\
			\hline
			\multirow{4}{*}{5~K}
			& 0   & 100 & 0  \\
			& 0.5 & 100 & 0  \\
			& 2.5 & 91   & 9 \\
			& 6   & 11  & 89 \\
			\hline
			\multirow{4}{*}{35~K}
			& 0   & 100 & 0   \\
			& 0.5 & 95  & 5   \\
			& 2.5 & 36  & 64  \\
			& 6   & 1   & 99 \\
			\hline
			\multirow{4}{*}{45~K}
			& 0   & 100 & 0  \\
			& 0.5 & 67  & 33 \\
			& 2.5 & 4   & 96 \\
			& 6   & 0   & 100 \\
		\end{tabular}
	\end{ruledtabular}
\end{table}

\section{DFT energy mapping and exchange interactions}
\label{sec:dft_mapping}

\subsection*{CrPS$_4$ Hamiltonian from DFT energy mapping}

\begin{figure*}[htb]
\includegraphics[width=\textwidth]{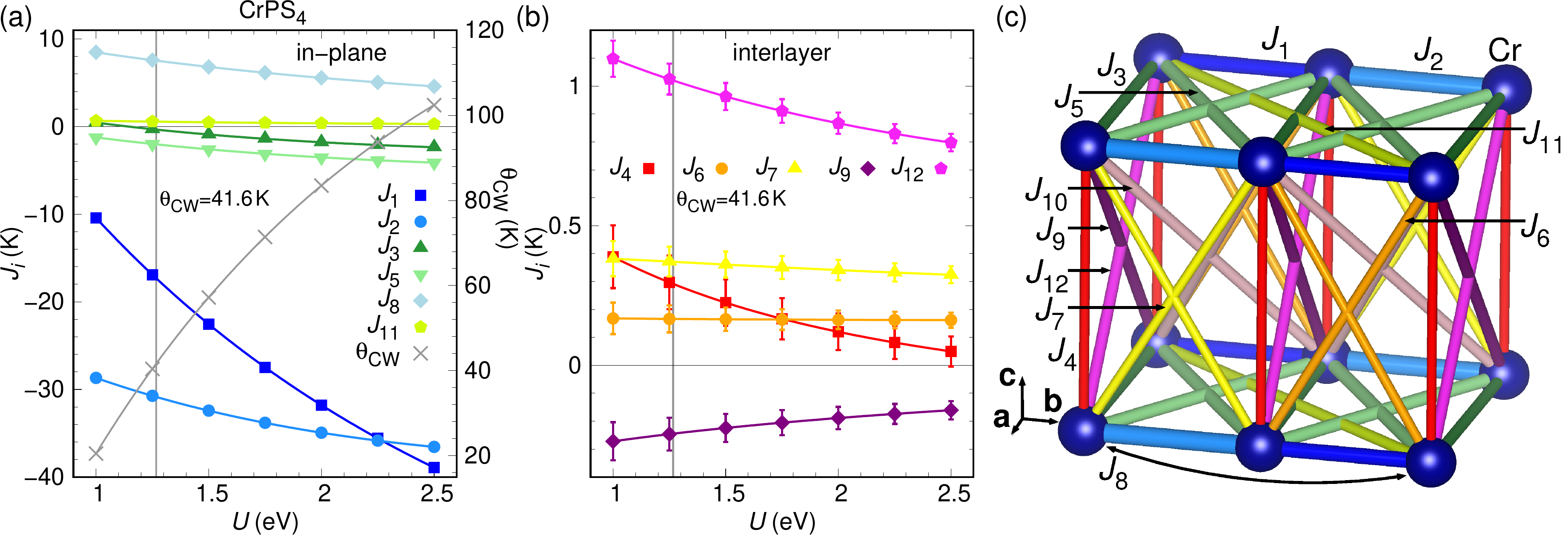}
\caption{Exchange couplings of CrPS$_4$, calculated within GGA+U at $J_H=0.72$~eV and $6\times 6\times 6$ $k$ points in a supercell containing 24 Cr sites. The vertical line indicates the $U$ value where the exchange couplings match the experimental Curie-Weiss temperature~\cite{CPS_Peng}.
}\label{fig:exchange}
\end{figure*}

\begin{figure}[htb]
\includegraphics[width=\columnwidth]{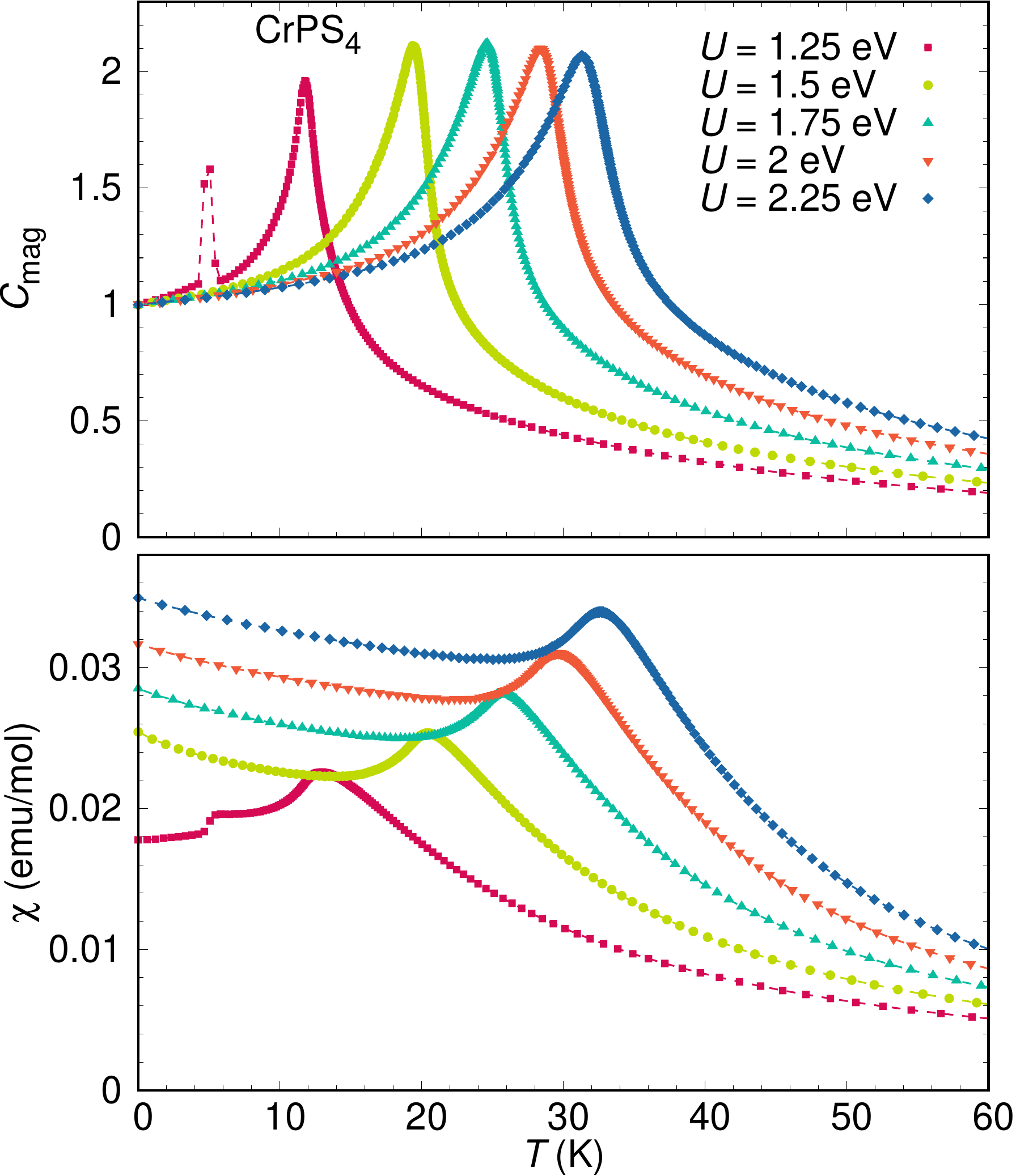}
\caption{Classical Monte Carlo simulation of the specific heat of CrPS$_4$, calculated with five of the parameter sets shown in Fig.~\protect\ref{fig:exchange}.}\label{fig:Cmag}
\end{figure}

\begin{figure}[htb]
\includegraphics[width=\columnwidth]{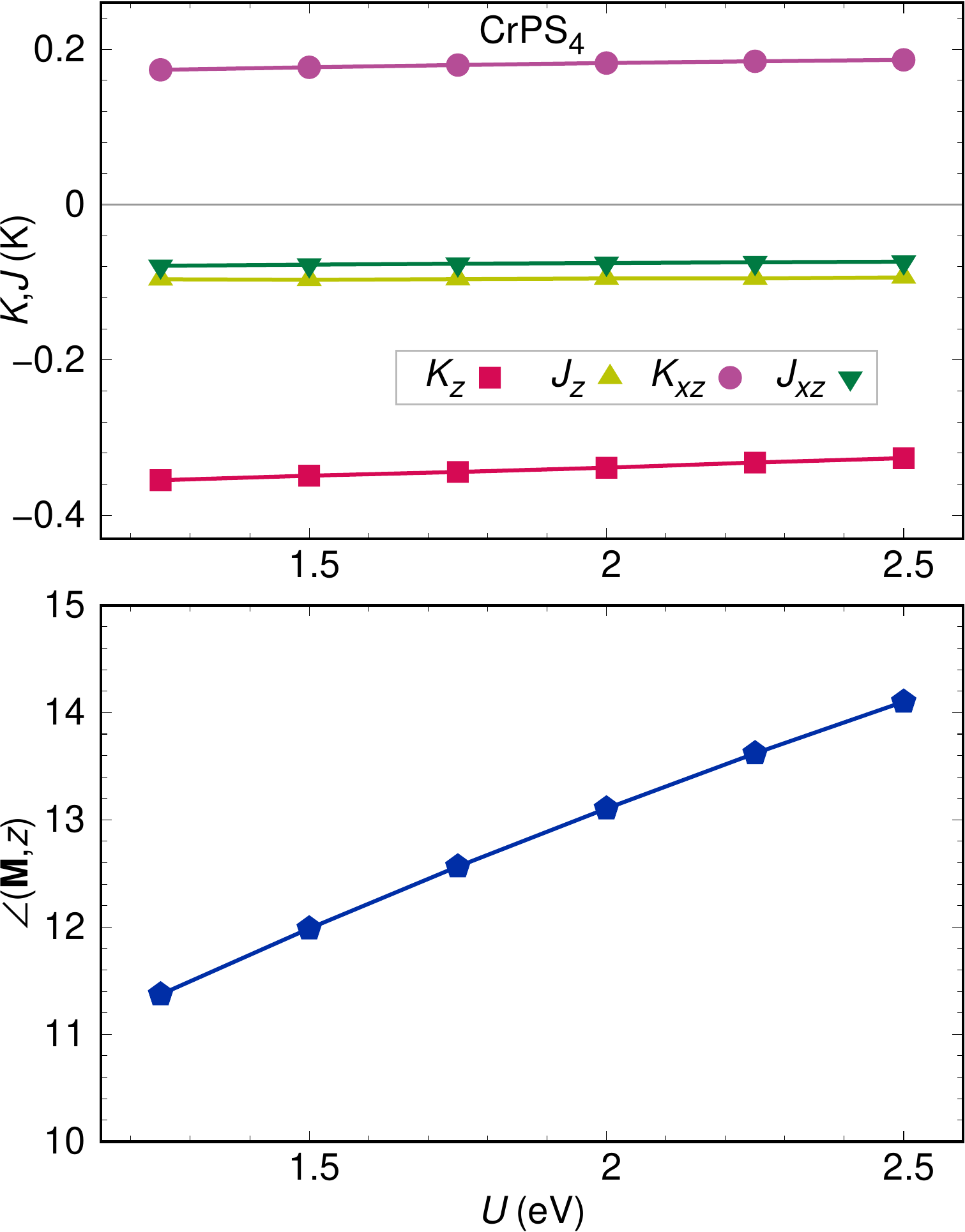}
\caption{(a) Anisotropic interactions of CrPS$_4$ as a function of on-site Coulomb interaction strength $U$. (b) Equilibrium tilt angle of the magnetic moments in the $ac$ plane for ferromagnetically ordered chains along $b$, measured from the $z$ direction.}\label{fig:aniso}
\end{figure}

CrPS$_4$ is dominated by one-dimensional magnetic exchange interactions, with two large couplings forming Cr chains along the $b$ direction~\cite{calder2020mag}. As one-dimensional magnets cannot order magnetically, the explanation of the N{\'e}el temperature of $T_\mathrm{N}=37$\,K requires careful evaluations of (i) the exchange interactions that connect the Cr chains in the $ab$ plane, (ii) the exchange that connects the planes along $c$ and (iii) the anisotropic interactions and in particular the single-ion anisotropy. 
We use density functional theory based energy mapping~\cite{Guterding2016,Iqbal2017} to determine the Heisenberg Hamiltonian parameters. For this purpose, we perform all electron density functional theory calculations using
the full potential local orbital (FPLO) code~\cite{Koepernik1999} in combination with the
generalized gradient approximation (GGA) exchange and correlation
functional~\cite{Perdew1996}. For the electronic structure calculations, we use the $T=60$~K crystal structure with $C2/m$ space group. We use a $\sqrt{4}\times 3\times \sqrt{4}$ supercell of CrPS$_4$. This allows us to determine sixteen exchange interactions, including many important in-plane and interlayer interactions. We deal with strong electronic correlations on the Cr $3d$ orbitals with a GGA+U exchange correlation functional~\cite{Liechtenstein1995} where we fix $J_H=0.72$~eV following Ref.~\onlinecite{Mizokawa1996}. We perform the calculations for five different values of $U$ and fit the results to the Heisenberg Hamiltonian
$$H=\sum_{i<j} J_{ij} {\bf S}_i\cdot {\bf S}_j$$
where spin operators ${\bf S}_i$ have length $3/2$ and bonds are not double-counted. The results are shown in Figs.~\ref{fig:exchange}\,(a) and (b). Fig.~\ref{fig:exchange}\,(c)
shows the chromium sublattice of CrPS$_4$
with bonds indicating the exchange pathways. 
We find two large ferromagnetic exchange interactions $J_1$ and $J_2$, forming 1D ferromagnetic chains along $b$, in agreement with the literature~\cite{calder2020mag}. The bonds $J_3$, $J_5$ and $J_{11}$ connect the FM chain in the $ab$ plane and are predominantly ferromagnetic as well (Fig.~\ref{fig:exchange}\,(a)). This suggests ferromagnetic order in the $ab$ plane. The planes are connected by a large number of interlayer couplings (Fig.~\ref{fig:exchange}\,(b)) which are predominantly antiferromagnetic, suggesting an A-type antiferromagnetic order of the 3D crystal.
The $U$ value that best describes CrPS$_4$ now has to be determined. One way would be to use the experimental Curie-Weiss temperature, for example $\theta_{\mathrm{CW}}=41.6$\,K~\cite{CPS_Peng} (see vertical line in Fig.~\ref{fig:exchange}\,(a) and (b)). However, the scale of CrPS$_4$ interactions has also been determined by Calder {\it et al.} from fits to inelastic neutron scattering dispersions which led to an optimal set $J_1=-34.4$\,K, $J_2=-25.3$\,K, $J_3=-5.9$\,K and $J_c\equiv J_4=1.9$\,K, this translates into $\theta_{\mathrm{CW}}=83$\,K, significantly higher than all values extracted by inverse susceptibility fits. In fact, this is quite reasonable because a Curie-Weiss temperature of 42\,K and an ordering temperature of $T_\mathrm{N}=37$\,K are difficult to reconcile in a predominantly 1D compound. Therefore, we keep an open mind and use classical Monte Carlo simulations for five of the Hamiltonians determined by DFT energy mapping at $U$ values of 1.25, 1.5, 1.75, 2 and 2.25\,eV, they correspond to Curie-Weiss temperatures $\theta_{\mathrm{CW}}=40$, 57, 72, 84 and 94\,K, respectively (see Fig.~\ref{fig:exchange}\,(a)).
The calculated specific heat for these five Hamiltonians is shown in Fig.~\ref{fig:Cmag}. Remarkably, the $U=2$\,eV Hamiltonian yields an ordering temperature of approximately 35\,K in classical Monte Carlo simulations, in close agreement with the experimental value of $T_\mathrm{N}=37$\,K and has the same energy scale as the Hamiltonian from INS~\cite{calder2020mag}.

\subsection{Additional DFT results}

\begin{figure}[htb]
\includegraphics[width=0.9\columnwidth]{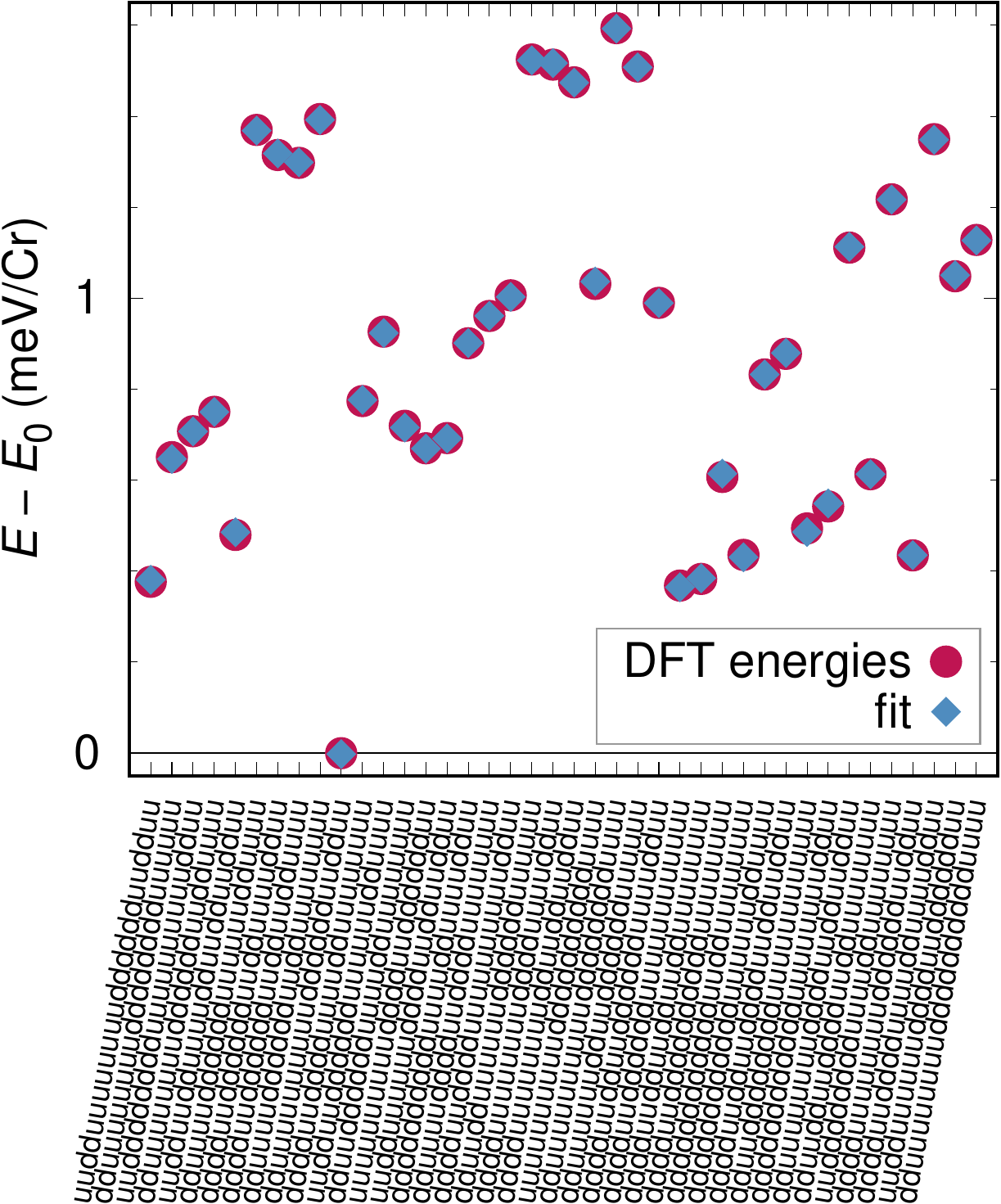}
\caption{Total energies of 40 different spin configurations of the  $\sqrt{4}\times 3\times \sqrt{4}$ supercell of  CrPS$_4$, calculated with GGA+U at $U=1.25$~eV (red symbols) and energies of the fitted Hamiltonian (blue symbols). The fit is excellent.}\label{fig:config}
\end{figure}

An example for the quality of the fit of the DFT energy mapping is given in Fig.~\ref{fig:config}.  

\section{Symmetry analysis and reduced Hamiltonian}
\label{sec:symmetry_ham}

In the following we concentrate on bonds 1 and 2, assuming that both Heisenberg and non-Heisenberg interactions on these bonds are much stronger than on the interchain ones. These bonds are similar, their lengths are 3.70 and 3.55~\AA, and the Cr--S--Cr angles are $94^\circ$ and $97^\circ$. Moreover, neither of these bonds allows Dzyaloshinskii--Moriya interactions, at least not within a single chain.

We now write the general form of the interaction $(i=1,2)$ in an arbitrary Cartesian coordinate system:
\begin{equation}
E_i = \sum_{\alpha\beta} S_\alpha U^{(i)}_{\alpha\beta} S'_\beta + \sum_{\alpha\beta} S_\alpha K^{(i)}_{\alpha\beta} S_\beta .
\label{eq:igor-general}
\end{equation}
The two spins at the ends of the bond are $S$ and $S'$. In principle, the following terms are allowed by symmetry:
\begin{equation}
K =
\begin{pmatrix}
K_{11} & K_{12} & K_{13}\\
K_{21} & K_{22} & K_{23}\\
K_{31} & K_{32} & K_{33}
\end{pmatrix}.
\end{equation}
Further, since $S_1^2+S_2^2+S_3^2=S^2=\mathrm{const}$, one can always rewrite the diagonal sums as
$K_1(S_1^2+S_2^2+S_3^2)+(K_2-K_1)S_2^2+(K_3-K_1)S_3^2$, that is, up to an unimportant isotropic term one can always set $K_{11}=0$. Now,
\begin{equation}
U =
\begin{pmatrix}
J_{11} & J_{12} & J_{13}\\
J_{21} & J_{22} & J_{23}\\
J_{31} & J_{32} & J_{33}
\end{pmatrix}.
\end{equation}
Similarly, one can always absorb $J_{11}$ into the isotropic Heisenberg term, setting $J_{11}=0$. From the fact that the bond center is an inversion center, the matrix is symmetric, so $K_{\alpha\beta}=K_{\beta\alpha}$ and $J_{\alpha\beta}=J_{\beta\alpha}$.

At this stage, we introduce several simplifying approximations. First, we set the monoclinic angle of $91.9^\circ$ to $90^\circ$, which is only a minor change. Second, we neglect interchain interactions and consider a single individual CrPS$_4$ chain. In addition to the inversion center, there is an $M_b$ mirror plane passing through $c$ and a $180^\circ$ rotation around $b$. Recalling the transformation properties of axial vectors under inversion, mirror, and twofold rotation,
\begin{center}
\begin{tabular}{c c c c}
 & $S_\alpha$ & $S_\beta$ & $S_\gamma$\\
$I$ & $S_\alpha$ & $S_\beta$ & $S_\gamma$\\
$M_\gamma$ & $-S_\alpha$ & $-S_\beta$ & $S_\gamma$\\
$C_{2\gamma}$ & $-S_\alpha$ & $-S_\beta$ & $S_\gamma$
\end{tabular}
\end{center}
where $\alpha,\beta,\gamma$ are Cartesian indices, one sees that under either $M_b$ or $C_{2b}$, $S_a$ and $S_c$ change sign while $S_b$ does not. This shows that the nondiagonal elements $K(J)_{ab}=K(J)_{bc}=0$. At the same time, $K(J)_{ac}$ can be nonzero, although it is expected to be small in magnitude.

In practice, the CrS$_6$ octahedra are rotated with respect to the crystallographic axes. For the purpose of setting an appropriate coordinate system, one may temporarily neglect the $\pm5^\circ$ wobbling caused by the phosphorus atoms and adopt a quasi-orthogonal Cartesian frame. Choosing $y\parallel b$, $x\parallel a$, and $z\approx c$, the following Hamiltonian emerges:
\begin{equation}
	\begin{aligned}
		\mathcal{H}_{ij} ={}&
		J\,\mathbf{S}\!\cdot\!\mathbf{S}'
		+ J_x S_x S'_x
		+ J_y S_y S'_y \\
		&+ J_{xz}
		\left(S_x S'_z + S_z S'_x\right) \\
		&+ K_y\left(S_y^2 + S_y'^2\right)
		+ K_z\left(S_z^2 + S_z'^2\right) \\
		&+ K_{xz}
		\left(S_x S_z + S'_x S'_z\right) .
		\label{eq:igor-main-ham}
	\end{aligned}
\end{equation}
This exhausts the onsite and nearest-neighbor magnetic interactions relevant for the reduced description used in the main text.

Accounting for the phosphorus-induced symmetry breaking, additional compass terms can be added:
\begin{equation}
J_{xy}(S_xS'_y+S_yS'_x)+K_{xy}(S_xS_y+S'_xS'_y).
\label{eq:igor-compass}
\end{equation}

\section{Spin-reorientation transition model}
\label{sec:srt-model}

In the simplified model, the total energy in the $xz$ plane can be written as
\begin{equation}
E = K_x M_x^2 + K_z M_z^2 + J_x\langle M_xM'_x\rangle + J_z\langle M_zM'_z\rangle .
\label{eq:igor-srt-1}
\end{equation}
Here $M_\alpha^2$ denotes the squared projection of a single spin, averaged over thermal fluctuations, while $\langle M_\alpha M'_\alpha\rangle$ is the corresponding spin--spin correlation function. The two quantities are identical only if the entire chain fluctuates coherently while remaining collinear. In practice, neighboring spins also fluctuate relative to one another, so $\langle M_\alpha M'_\alpha\rangle$ decays faster than $M_\alpha^2$.

This can be approximately incorporated by rewriting Eq.~(\ref{eq:igor-srt-1}) as
\begin{equation}
E = \tilde K_x M_x^2 + \tilde K_z M_z^2 + \xi(T)\,\tilde J_x M_xM'_x + \xi(T)\,\tilde J_z M_zM'_z ,
\label{eq:igor-srt-2}
\end{equation}
where $\xi(T)$ depends on temperature roughly as $\langle M_\alpha M'_\alpha\rangle/M_\alpha^2$, varying from $\xi\approx 1$ at $T=0$ to $\xi\ll 1$ near $T\approx T_\mathrm{N}$. This simplified Hamiltonian has only discrete easy-axis solutions.

In the full model relevant to the observed rotation in the $ac$ plane,
\begin{equation}
	\begin{aligned}
		E ={}&
		K_x M_x^2
		+ K_z M_z^2
		+ 2K_{xz} M_x M_z \\
		&+ J_x \langle M_x M'_x\rangle
		+ J_z \langle M_z M'_z\rangle \\
		&+ J_{xz}
		\bigl(
		\langle M_x M'_z\rangle
		+
		\langle M_z M'_x\rangle
		\bigr) .
		\label{eq:igor-srt-3}
	\end{aligned}
\end{equation}
The spins can rotate continuously in the $xz$ plane according to the principal values of the tensor
\begin{equation}
\begin{pmatrix}
K_x+\xi\tilde J_x & K_{xz}+\xi J_{xz}\\
K_{xz}+\xi J_{xz} & K_z+\xi\tilde J_z
\end{pmatrix}.
\label{eq:igor-srt-matrix}
\end{equation}

\section{Temperature-dependent correlation-renormalization estimator}
\label{sec:correlations}

We give here a simple phenomenological estimate for the ordered-state correlation-renormalization factor used in the main text. The goal is not to construct a microscopic theory of the paramagnetic regime, but to estimate how the intersite spin correlation is suppressed relative to the on-site spin fluctuation as the ordered moment decreases.

Assume that the spin fluctuates by a small angle $\varphi$ away from the local easy axis, with $\varphi$ Gaussian distributed. For fluctuations in the $ac$ plane, the relevant on-site and intersite averages are
\begin{equation}
\langle M_z^2\rangle = \mu_2/\mu_0,
\qquad
\langle M_zM'_z\rangle = \langle M_z\rangle^2 = (\mu_1/\mu_0)^2,
\end{equation}
where
\begin{equation}
\mu_i = \int \cos^i\varphi \, \exp(-\varphi^2/\delta^2)\, d\varphi ,
\end{equation}
and $\delta$ is the fluctuation amplitude. This gives
\begin{equation}
\langle M_z\rangle = e^{-\delta^2/4},
\qquad
\langle M_zM'_z\rangle = e^{-\delta^2/2}
= \langle M_z\rangle^2,
\end{equation}
while
\begin{equation}
\langle M_z^2\rangle =
\frac{1+e^{-\delta^2}}{2}
= \frac{1+\langle M_z\rangle^4}{2}.
\end{equation}
Therefore,
\begin{equation}
\xi(T) \equiv
\frac{\langle M_zM'_z\rangle}{\langle M_z^2\rangle}
=
\frac{2\langle M_z\rangle^2}{1+\langle M_z\rangle^4}.
\label{eq:igor-xi}
\end{equation}
Identifying $m(T)=\langle M_z\rangle/M(0)$ gives the expression used in the
main text,
\begin{equation}
\xi(T)=\frac{2m(T)^2}{1+m(T)^4}.
\end{equation}
This expression should be regarded as a phenomenological ordered-state estimator for the relative suppression of intersite correlations compared with on-site fluctuations. It is not intended as a microscopic description of the paramagnetic regime, where the mPDF-derived nearest-neighbor correlation is used directly.

\bibliography{ref}